\def\year{2022}\relax
\documentclass[letterpaper]{article} 
\usepackage{aaai21}  
\usepackage{times}  
\usepackage{helvet} 
\usepackage{courier}  
\usepackage[hyphens]{url}  
\usepackage{graphicx} 
\usepackage{natbib}  
\usepackage{caption} 
\usepackage{booktabs}
\usepackage[table]{xcolor}
\usepackage{array} 
\usepackage[hyphens]{xurl}
\usepackage[hidelinks]{hyperref}
\usepackage{multirow}
\usepackage{graphicx,subcaption}
\usepackage{siunitx}
\usepackage{booktabs}

\newcounter{KNNumberOfComments}
\stepcounter{KNNumberOfComments}

\newcounter{JSNumberOfComments}
\stepcounter{JSNumberOfComments}

\newcommand{\bm}[1]{\textcolor{black}{#1}}

\title{``This is Fake News'': Characterizing the Spontaneous Debunking from \\Twitter Users to COVID-19 False Information}

\author{
Kunihiro Miyazaki\textsuperscript{\rm 1},
Takayuki Uchiba\textsuperscript{\rm 2},
Kenji Tanaka\textsuperscript{\rm 1},\\
Jisun An\textsuperscript{\rm 3},
Haewoon Kwak\textsuperscript{\rm 3},
Kazutoshi Sasahara\textsuperscript{\rm 4}\\
}
\affiliations{
\textsuperscript{\rm 1 }The University of Tokyo, Japan\\
\textsuperscript{\rm 2 }Sugakubunka, Japan\\
\textsuperscript{\rm 3 }Singapore Management University, Singapore\\
\textsuperscript{\rm 4 }Tokyo Institute of Technology, Japan\\
\{kunihirom,jisun.an,haewoon\}@acm.org, takayuki.uchiba@sugakubunka.com, \\
tanaka@tmi.t.u-tokyo.ac.jp, sasahara.k.aa@m.titech.ac.jp
}

\begin{document}

\maketitle

\begin{abstract}
False information spreads on social media, and fact-checking is a potential countermeasure.
However, there is a severe shortage of fact-checkers; an efficient way to scale fact-checking is desperately needed, especially in pandemics like COVID-19.
In this study, we focus on spontaneous debunking by social media users, which has been missed in existing research despite its indicated usefulness for fact-checking and countering false information.
Specifically, we characterize the tweets with false information, or fake tweets, that tend to be debunked and Twitter users who often debunk fake tweets.
For this analysis, we create a comprehensive dataset of responses to fake tweets, annotate a subset of them, and build a classification model for detecting debunking behaviors.
We find that most fake tweets are left undebunked, spontaneous debunking is slower than other forms of responses, and spontaneous debunking exhibits partisanship in political topics. 
These results provide actionable insights into utilizing spontaneous debunking to scale conventional fact-checking, thereby supplementing existing research from a new perspective.
\end{abstract}

\section{Introduction}

The spread of false information has been a severe problem in our society~\cite{lazer2018science}.
In the recent COVID-19 pandemic, false information and its spread has been considered as dangerous as the virus~\cite{guardian2021}. 
For example, the information about wrong treatment has led people to die~\cite{Hundreds43:online}, and a conspiracy theory about vaccines has made people less likely to get vaccinated, which unnecessarily undermines the utility of society as a whole~\cite{burki2019vaccine}.
The World Health Organization (WHO) called the prevalence of such false information an ``infodemic''~\cite{Managing84:online} and has set it as an important global issue.

Various countermeasures have been implemented in order to combat this infodemic, among which fact-checking is a prominent approach.
Fact-checking is an activity to verify the correctness of the information, news, and discourse spread~\cite{walter2020fact}, which is conducted by an individual, a group, or an organization, e.g., Snopes\footnote{\url{https://www.snopes.com/}} and Vaccination Demand Observatory\footnote{\url{https://vaccinationdemandobservatory.org/}}.
However, it takes time and resources to train fact-checkers~\cite{graves2017anatomy}. 
Given the volume of false information generated and circulated on social media~\cite{America18:online}, 
it is almost impossible to have enough fact-checkers to verify all the suspicious information~\cite{Criticsf50:online}.

Academics and the industry have been exploring ways to scale fact-checking.
The efforts can be divided into largely two directions: automated fact-checking~\citep[e.g.,][]{liu2015real} and fact-checking by crowdsourcing~\citep[e.g.,][]{allen2020scaling}.
Automated fact-checking mainly relies on algorithms such as machine learning to detect false information and has already contributed to removing fake news from some platforms ~\cite{Facebook94:online}.
Fact-checking by crowdsourcing exploits the wisdom of crowds. It is reported that their judgment has a high correlation with professional judgment in fact-checking~\cite{epstein2020will}, although not yet as accurate as professionals~\cite{godel2021moderating}.
While these methods are promising to combat the infodemic, machine-learning-based models are known to be highly context-dependent~\cite{bang2021model}; they tend to perform poorly for newly circulating false information, and crowdsourcing requires a significant amount of time and cost if the crowd needs to verify a large amount of information.

\begin{figure}[!t]
\centering
  \begin{subfigure}[t]{.56\linewidth}
    \centering
    \includegraphics[width=\linewidth]{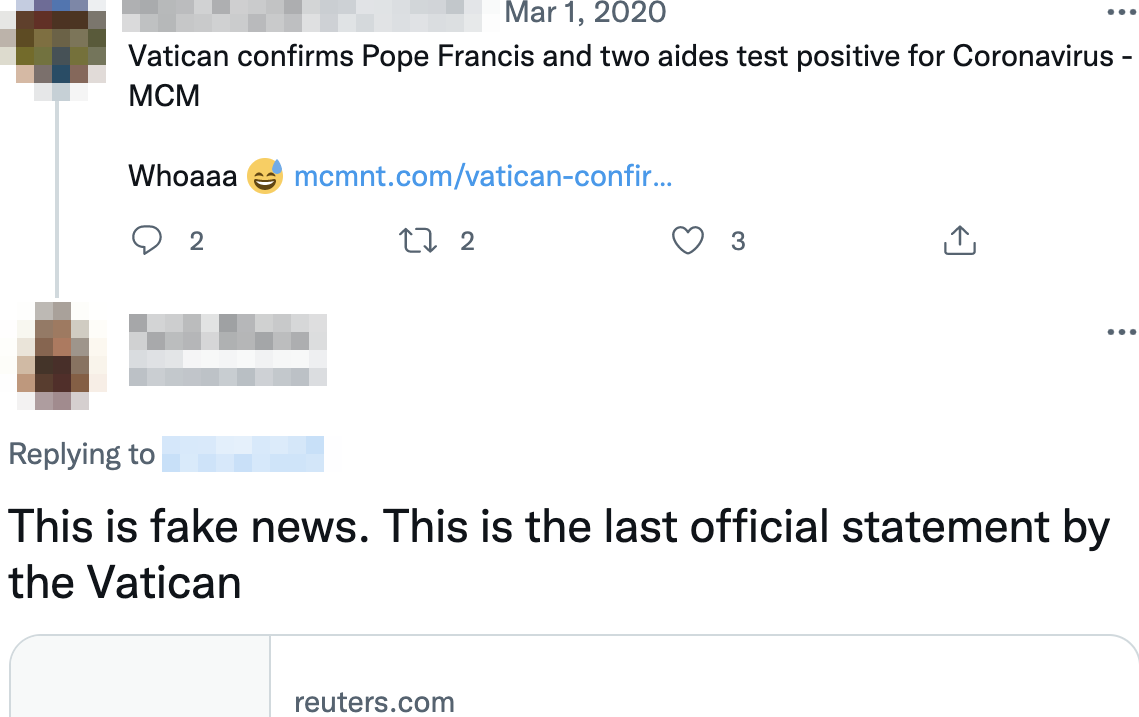}
    \caption{Reply.}
    \label{fig:1a_sb}
  \end{subfigure}
  \begin{subfigure}[t]{.43\linewidth}
    \centering
    \includegraphics[width=\linewidth]{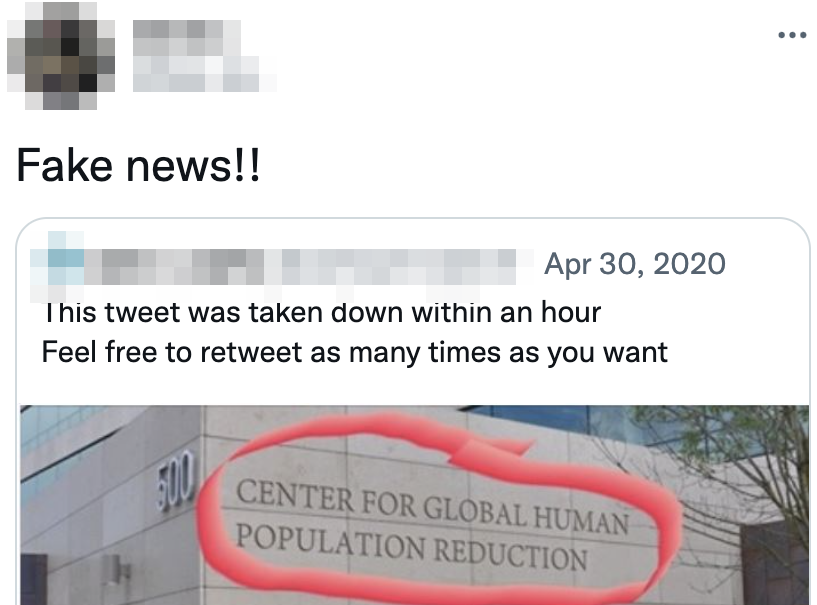}
    \caption{Quote Tweet (QT).}
    \label{fig:1b_sb}
  \end{subfigure}
  \caption{Examples of spontaneous debunking.}
  \label{fig:examples_of_spontaneous_debunking} 
\end{figure}

Spontaneous debunking, one type of debunking by crowds, is a user's act of pointing out false information in other users' posts on social media (i.e., social correction~\cite{bode2018see}) without any particular incentive, which we can often see in the wild.
Figure~\ref{fig:examples_of_spontaneous_debunking} shows examples of spontaneous debunking on Twitter: 
Twitter users voluntarily pointed out that the original tweet contains false information by reply and retweet with comments (i.e., Quote tweet, QT).
Understanding the characteristics of these spontaneous debunking actions allows various ways to leverage them.
First, it may be possible to prioritize potential false information that needs verification by observing the amount and tendency of spontaneous debunking toward them. 
Second, by understanding their motivation, it may be possible to encourage more participation in debunking. 
Third, spontaneous debunkers can work as social sensors~\cite{liu2015social} to detect false information quickly. 
Despite having great potential, spontaneous debunking has been less explored.

With this in mind, we aim to characterize spontaneous debunking on social media.
We exhaustively collect tweets containing false information and responses (i.e., reply and QT) from other users toward these tweets.
Then, we identify \textit{debunking} tweets among the responses by using the language model fine-tuned by our annotated dataset.
After a first glance at our dataset, we pose the following research questions (RQs) to examine tweets with false information and debunking:
\begin{itemize}
    \item RQ1: What are the characteristics of tweets with false information (fake tweets) that tend to be debunked? 
    \item RQ2: What are the characteristics of spontaneous debunkers?
\end{itemize}

Our analysis confirms that much of the fake news is left undebunked and debunking behavior is generally slower than other responses.
Fake tweets related to the factual tweets, such as the status of infection around the world and the countermeasures against COVID-19, are especially less debunked than other topics. 
We also find that the most frequent debunkers are highly partisan and well connected in each group.

Our contributions are as follows. We create a spontaneous debunking dataset consisting of tweets with false information and the responses to them. 
We annotate 10,000 responses, including debunking and non-debunking behavior. 
We build the sentence classification model by the annotated samples and classify all the responses. 
The codes and data would be available upon publication. Our dataset will be a valuable resource for upcoming studies. 
We shed light on debunking behavior by characterizing fake tweets likely to be debunked, debunking tweets, and debunkers, which has been underexplored by previous research.

\section{Related Works}

\subsection{Social correction}

Debunking is an action to inform authors of posts as inaccurate, which is reported to generally reduce false beliefs of the authors~\cite{wood2019elusive}.
To debunk as much false information as possible is important because leaving false information undebunked is said to raise the ``implied truth effect,''~\cite{pennycook2020implied} where undebunked false information gives its witnesses the impression that it is true.
On the other hand, a potential ``backfire effect,'' where being debunked would ironically counter-strengthen its originator's beliefs, has been debated for a long time~\citep[e.g.,][]{lewandowsky2012misinformation}. 
Nonetheless, recent research suggests that the support for this effect is small~\citep[e.g.,][]{swire2020searching}.

Debunking by social media users is referred to as social correction~\cite{kligler2021collective}.
It is known that social correction is effective not only in curbing false perceptions of authors of fake posts~\cite{shu2020fakenewsnet} but also in reducing the false belief of those who witness false information being debunked on social media~\cite{bode2018see,colliander2019fake}.
In addition, debunking by crowdsourcing showed that the crowd's assessment of the credibility of the news publisher (e.g., publisher's website domain) was highly correlated with the assessment of professional fact-checkers, which showed the effectiveness of the wisdom of crowds, however unstable~\citep[e.g.,][]{pennycook2019fighting}.
Moreover, in the literature on automated fake news detection, social contexts, such as responses from other users, are known to be essential signals to significantly improve the accuracy of the model~\citep[e.g.,][]{cui2019same}.
Nevertheless, the characteristics of debunking behavior from social media users have not been fully explored. 

In this work, we call the debunking behavior ``spontaneous debunking'' as we focus particularly on the social correction without any incentives, which is different from artificially-generated debunking in lab/field experiments~\cite{mosleh2021perverse}.
The study by~\citet{vo2018rise}, which analyzed how social media users use fact-checked information, e.g., by Snopes, especially in replies, is the closest to our work. 
We also analyze replies, but the difference is that we ensure the original posts targeted by replies contain false information. 
This difference comes from the different purposes of the analysis. 
They ultimately focused on how the fact-checked information is used and shared, whereas our primary focus is on which fake tweets are most likely to attract debunking, which necessitates that the targeted tweets always contain false information.
Additionally, we analyze ``other responses'' that are not debunking. By basing our analysis on comparisons between debunking behavior and other responses, we can more clearly characterize debunking behavior (e.g., in the analysis of ``speed of debunking'').
Also, \citet{vo2019learning} examined the linguistic characteristics of tweets containing fact-checked information and attempted to create a model for automatically generating fact-checked tweets, which is different from our primary objective of analyzing the behavioral characteristics of debunking.

\subsection{Twitter Birdwatch}
Twitter has recently started a new initiative called ``Birdwatch''~\cite{Introduc78:online}.
It allows Twitter users to report suspicious tweets they encounter.
Thus, it can be called fact-checking that leverages the collective intelligence of social media. 
The data of reported suspicious tweets are publicly available and have already started being analyzed~\cite{prollochs2021community,allen2022birds}.

The difference between this study and Birdwatch is that Birdwatch collects the reporting from users, not publicly debunking the tweets that have false information.
In addition, the Birdwatch data contains tweets reported as false by Twitter users only.
This study analyzes both tweets with false information debunked by Twitter users \emph{and} those undebunked. 

\section{Building a Spontaneous Debunking Dataset}

We aim to analyze false information during COVID-19 and how it is debunked (and not debunked). 
In doing so, we first collect tweets that contain false information from various sources ($\S$3.1), and then we collect  `responses,’ i.e., replies and QTs ($\S$3.2).
We use the term \textit{fake tweet} to refer to a tweet containing false information, such as false claims and fake news.

\subsection{Collection of fake tweets}

The widely used methods to collect fake tweets are machine-learning-based inference~\cite{patwa2021overview}, hashtag search~\cite{al2018fake}, and getting neighbors of fake tweets using edit distance or distance in embedding space~\cite{shaar2020known}. 
However, none of these methods can guarantee high veracity~\citep[e.g.,][]{bang2021model}, and cannot avoid containing non-fake tweets.
We need to avoid false positives as much as possible since the primary focus of this work is \textit{debunking behavior} to fake tweets.
To this end, we take two approaches: domain-based and claim-based approaches, which can achieve relatively high veracity in collecting fake tweets.
In both, we leverage already-fact-checked tweets and domains from previous research.

\noindent \textbf{Claim-based approach.} 
We utilize tweet datasets that were manually fact-checked and confirmed as fake tweets in existing studies.
We adopt the following conditions to select datasets:
(1) fake tweets, (2) in English, (3) about COVID-19, (4) labeled by the fact-checking organizations or experts, (5) whose labels are publicly available, and (6) whose IDs are also publicly available (e.g., the dataset of~\citet{patwa2021overview} provides text, but we need tweet IDs for the search of responses).
We examine publicly available datasets~\citep[see the survey of ][]{murayama2021dataset} to see if they satisfy the above conditions. 
This results in the six datasets shown in Table~\ref{table:datasets_prev}.
Some datasets have fewer tweets than the original because we only use tweets labeled as fake tweets.

\begin{table}[!ht]
\small
    \centering
\begin{tabular}{p{6.5cm}c}
\toprule
Datasets                                                                   & Count \\
\midrule
CoAID~\cite{cui2020coaid}                           & 7,267  \\
FibVID~\cite{kim2021fibvid}                        & 743   \\
COVIDLies~\cite{hossain2020covidlies}               & 114   \\
COVID-Alam~\cite{alam2021fighting}                  & 6     \\
CMU-MisCov19~\cite{memon2020characterizing}         & 855   \\ 
Misinformation\_COVID19~\cite{shahi2021exploratory} & 1,221  \\
\bottomrule
\end{tabular}
\caption{\label{table:datasets_prev}
Datasets that were manually fact-checked in existing studies and the number of tweets labeled as fake tweets.
}
\end{table}

These tweets total 18,929.
We retrieve these tweets using their IDs and get 10,190 tweets (Table~\ref{table:fake}). 
The rest of the 8,738 tweets have already been deleted or their accounts are no longer public as of November 2021. 

\noindent \textbf{Domain-based approach.}
We also collect fake tweets by searching for tweets containing URLs of suspicious domains in our COVID-19-related tweet data set.
Our COVID-19-related tweet dataset consists of 86,357,693 English-language tweets collected from February 2020 to February 2021 excluding retweets. %
The query words for building this dataset are ``corona virus,'' ``coronavirus,'' ``COVID19,'' ``2019-nCoV,'' ``SARS-CoV-2,'' and ``wuhanpneumonia.'' 
The suspicious domains are from the list from CoVaxxy~\cite{deverna2021covaxxy}, which is the dashboard of COVID-19 misinformation on social media.
This list is publicly available and consists of 674 domains.
We obtain the fake tweets by searching all domains in this list.
As a result, we retrieved 333,470 tweets with URLs  (Table~\ref{table:fake}).

\begin{table}[!ht]
\small
    \centering
    \scalebox{0.92}{
    \begin{tabular}{lllcc}
    \toprule
    Approach & \multicolumn{2}{c}{Type of tweets}                             & Count   & Ratio (\%) \\ 
    \midrule
    \multirow{5}{*}{Claim-based}  & \multicolumn{2}{c}{All}               & 10,190  & 100          \\ \cmidrule{2-5} 
                                  & \multirow{2}{*}{Responded} & Reply & 2,200   & 21.59      \\
                                  &                               & QT    & 1,885   & 18.50      \\ \cmidrule{2-5} 
                                  & \multirow{2}{*}{Debunked} & Reply & 1,521   & 14.93      \\
                                  &                               & QT    & 1,184   & 11.62      \\ \midrule
    \multirow{5}{*}{Domain-based} & \multicolumn{2}{c}{All}               & 333,470 & 100          \\ \cmidrule{2-5} 
                                  & \multirow{2}{*}{Responded via} & Reply & 30,296  & 9.09       \\
                                  &                               & QT    & 17,166  & 5.15       \\ \cmidrule{2-5} 
                                  & \multirow{2}{*}{Debunked via} & Reply & 10,504  & 3.15       \\
                                  &                               & QT    & 3,429   & 1.03       \\ \bottomrule
    \end{tabular}
    }
\caption{\label{table:fake}
Statistics of fake tweets. 
The displayed numbers are the amount (and ratio) of all the fake tweets, those that get responses at least once and that is debunked at least once.
}
\end{table}

\subsection{Collection of Responses to Fake Tweets}
We collect replies and QTs as responses to fake tweets.

\noindent \textbf{Replies.}
We search fake tweet IDs and obtain replies using conversation\_id\footnote{\url{https://developer.twitter.com/en/docs/twitter-api/conversation-id}}.
We get 2,639,104 replies.
We then remove non-English replies (25.7\% of the total).
Also, as the search with conversation\_id returns all reply trees, including replies to a reply, we only keep the direct replies to the fake tweets and eliminate further replies.
Finally, we get 1,249,882 replies in total (1,011,179 for claim-based and 318,154 for domain-based fake tweets, see Table~\ref{table:response}).

\noindent \textbf{Quote tweets.}
We obtain QTs by searching with the condition that the URL included in the tweet contains the ID of the targeted tweet because Twitter's API treats the quoted tweets as URLs, just like news media.
We obtain 884,060 QTs, and retain only those in English (34.8\% of total).
As a result, we get 527,007 QTs in total (470,349 for claim-based and 56,658 for domain-based fake tweets, see Table~\ref{table:response}).

\begin{table}[!ht]
\small
    \centering
    \scalebox{0.90}{
\begin{tabular}{llccc}
\toprule
Response               & \begin{tabular}[c]{@{}l@{}}Original \\      fake tweets\end{tabular} & Total     & Debunking & \begin{tabular}[c]{@{}l@{}}Ratio of\\      debunking\end{tabular} \\ \hline
\multirow{3}{*}{Reply} & Claim-based                                                          & 1,011,179 & 481,538   & 47.6\%                                                            \\
                       & Domain-based                                                         & 238,703   & 78,164    & 32.7\%                                                            \\ \cline{2-5} 
                       & Total                                                                & 1,249,882 & 559,702   & 44.8\%                                                            \\ \hline
\multirow{3}{*}{QT}    & Claim-based                                                          & 470,349   & 117,777   & 25.0\%                                                            \\
                       & Domain-based                                                         & 56,658    & 9,458     & 16.7\%                                                            \\ \cline{2-5} 
                       & Total                                                                & 527,007   & 127,235   & 24.1\%                                                            \\ \bottomrule
\end{tabular}
}
\caption{\label{table:response}
Statistics of responses to fake tweets. 
}
\end{table}

\section{Detection of Debunking Behavior}
We first identify debunking behaviors among the responses by using the sentence classification model. 
In this study, we build a machine learning model to classify whether or not the responses to fake tweets are debunking behavior, where the input is the text of the response and the target variable is the type of response.

\subsection{Annotating Debunking Tweets}

Since no ground truth labels for debunking tweets of COVID-19 false information exist, we manually label a subset of the responses to fake tweets using Amazon Mechanical Turk (MTurk).
We aim to annotate 5,000 replies and QTs each.
As the ratio of debunking in replies and QTs is unknown, we carefully design a multistep annotation process to avoid class imbalance, which often impacts the model performance~\cite{tayyar-madabushi-etal-2019-cost}.
We first randomly sample 3,000 replies and 3,000 QTs and annotate them. Then, we find that 47.5\% of replies and 25.2\% of the QTs are labeled as debunking. 
We randomly sample 2,000 more replies because replies do not show a significant class imbalance. 
As for QTs, we randomly sample 2,000 more from quasi-debunking QTs, which is the preliminary model's output fine-tuned by the first 3,000 QTs. 
The preliminary model is trained with the undersampling technique to address the imbalances in the first 3,000 QTs.

We assign three annotators for each tweet in MTurk. We choose workers who 1) report their location as the U.S., 2) mark their approval rate as greater than 98\%, and 3) have a number of approved tasks (HIT) greater than 10,000, referring to the official guideline of MTurk~\cite{mturk2019}.
Furthermore, we choose ``Master Workers'' whose quality of work is officially verified by MTruk~\cite{Mturk2015}.
Since \emph{debunking} may be an ambiguous word, we offer concrete guidelines and examples. 
We ask the annotators ``Do you think the displayed reply/quote tweet (retweet with comment) is critical of the original tweet and debunks it?''. We show examples to annotators as follows:
\begin{itemize}
    \item Debunking: Point out fakes, Insult to tweet author, Refute with logic, Order to retract: e.g., ``This is fake news,'' ``You are mad, liar,'' ``Ginger does not cure COVID-19,'' ``Retract.''
    \item Others (Support, Comment, Queries, etc.): e.g., ``True,'' ``How did it happen?'' ``I heard this news on yesterday.''
\end{itemize}

If the class is ambiguous, we ask the annotators to choose Others.
The categories and examples of debunking are iteratively refined through pilot tests. 
The examples of Others as Support, Comment, and Queries come from the previous study on rumor detection~\cite{cheng2020vroc,cheng2021covid}.
As a result, we get the labels with the Fleiss Kappa score at 0.542 for replies and 0.541 for QTs, which are moderate agreement~\cite{landis1977measurement}.
The majority voting of the three annotators decides the final label of each response. 
Table~\ref{table:annot} shows the summary of the annotation results.

\begin{table}[!ht]
\scalebox{0.87}{
\begin{tabular}{lccccccc}
\hline
          & \multicolumn{3}{c}{Reply}               &  & \multicolumn{3}{c}{QT}                 \\ \cline{2-4} \cline{6-8} 
          & \multicolumn{2}{c}{Agreement} &         &  & \multicolumn{2}{c}{Agreement} &        \\ \cline{2-3} \cline{6-7}
Labels    & 2             & 3             & Total   &  & 2             & 3             & Total  \\ \hline
Debunking & 785           &  1,629            &  2,414     &  &  884            &  1,544            &  2,428    \\
Others    &  934            &  1,652            &  2,586     &  &  837            &  1,735          &  2,572    \\ \hline
Total         &  1,719           &  3,281           &  5,000    &  &  1,721           &  3,279           & 5,000  \\ \hline
\end{tabular}
}
\caption{\label{table:annot}
Result of annotation by three annotators for Replies and QTs.
The agreement indicates the number of annotators whose responses matched, i.e., 3 indicates the complete agreement.
Labels are determined by majority vote, e.g., among the 5,000 replies, all three annotators agree 1,629 replies are debunking, while 785 replies are decided as debunking from voting with 2 Debunking vs. 1 Others.
}
\end{table}

\subsection{Building a Debunking Classifier}
We evaluate several choices to build the classification model.
First, we adopt the basic machine learning models, i.e., Logistic Regression (LR), Random Forest (RF), and Linear SVM (SVM).
As the input, we use embedding features by SentenceBERT~\cite{reimers2019sentence}, which has been popularly used to represent a short text~\cite{an2021predicting}.
We also use pre-trained BERT models for prediction by fine-tuning them with our annotated dataset. 
In particular, we use: regular BERT~\cite{devlin2019bert}, trained on Wikipedia data (BERT-Wiki); BERTweet~\cite{nguyen2020bertweet}, trained on the Twitter corpus of general topics (BERT--Tweet); and COVID-Twitter-BERT, trained on the Twitter corpus on COVID-19 topics (BERT--Tweet--COVID19)~\cite{muller2020covid}.
For evaluation, we obtain the average F1 score by the 10-fold cross-validation.

\begin{table}[!ht]
\centering
\small
\begin{tabular}{lcccc}
\hline
                          & \multicolumn{2}{c}{Reply} & \multicolumn{2}{c}{QT} \\ \cline{2-5} 
Model                     & Mean        & SD         & Mean       & SD       \\ \hline
SBERT + LR                & 0.703       & $\pm$0.023       & 0.735      & $\pm$0.019     \\
SBERT + RF                & 0.673       & $\pm$0.021       & 0.699      & $\pm$0.024     \\
SBERT + SVM               & 0.716       & $\pm$0.017       & 0.736      & $\pm$0.016     \\
BERT--Wiki              & 0.739       & $\pm$0.023       & 0.752      & $\pm$0.023     \\
BERT--Tweet            & \textbf{0.792}      & $\pm$0.018       & 0.783      & $\pm$0.029     \\
BERT--Tweet--COVID19 & \textbf{0.792}       & $\pm$0.018       & \textbf{0.820}      & $\pm$0.030     \\ \hline
\end{tabular}
\caption{F1 scores for prediction of debunking behaviors by the 10-fold cross-validation. The highest scores in Reply and QT are shown in bold.}
\label{fig:evaluate_pred} 
\end{table}

\bm{Table~\ref{fig:evaluate_pred} shows the performance of various models. 
We find that the BERT--Tweet--COVID19 model shows the best performance for both reply and QT---we obtain the F1 score of 0.792 (SD $=\pm$0.018) for replies and 0.820 (SD $=\pm$0.030) for QTs.
To further demonstrate the generality of our model in the full dataset, we conduct a manual evaluation. 
We randomly extract 100 replies and 100 QTs that are not used in MTurk annotation, and then manually annotate them whether they are actual debunking or not. Compared to the predicted results, we get F1 scores of 0.773 for replies and 0.813 for QTs. These values are almost the same as the result of the original 10-fold cross-validation, and thus the result demonstrates the generality of our model.
Also, as the classification model with an F1 score of about 0.8 has been employed in previous studies~\cite{he2021racism}, in this study, we use our BERT--Tweet--COVID19 model to analyze the overall tendency of a large amount of data (330k fake tweets and 1.8m responses) in the subsequent analyses.}

Finally, we fine-tune the model using all annotated samples and infer whether or not the collected replies and QTs are debunking.
As a result, we get 559,702 debunkings from replies (44.8\% of total) and 127,235 debunkings from QTs (24.1\% of total), which are summarized in Table~\ref{table:response}.

\section{A First Look at Debunking}
In this section, we conduct an exploratory analysis to provide an overview of debunking behavior.  

\noindent \textbf{Do fake tweets receive (debunking) responses?}
The majority of fake tweets do not get any responses.
As Table~\ref{table:fake} in $\S$3.1 shows, 21.59\% and 18.50\% of claim-based fake tweets get replies and QTs, respectively, and 9.09\% and 5.15\% of domain-based fake tweets get replies and QTs, respectively.
Among them, the tweets debunked by replies and QTs are even lower in number. 
In the same table, we can see the proportion dramatically drops to 1.03\% (domain-based tweets debunked by QTs).

\begin{figure}[!h]
\centering
\includegraphics[width=\linewidth]{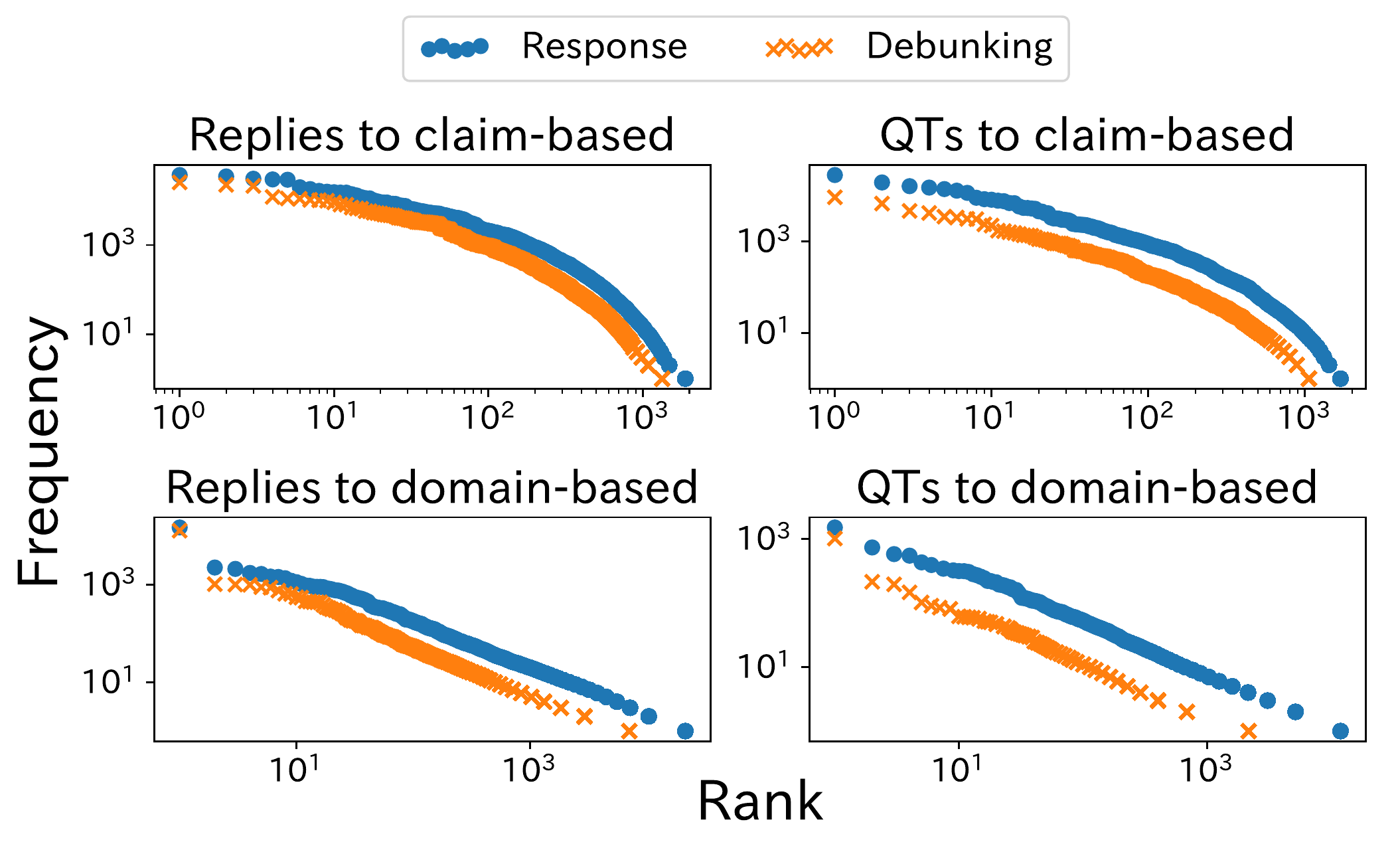}
\caption{Log-log plots for fake tweets in terms of frequency of responses/debunking to fake tweets (x-axis) and their rank of frequency (y-axis).}
\label{fig:zipf} 

\end{figure}

\begin{figure}[!ht]
\centering
  \begin{subfigure}[h]{.49\linewidth}
    \centering
    \includegraphics[width=\linewidth]{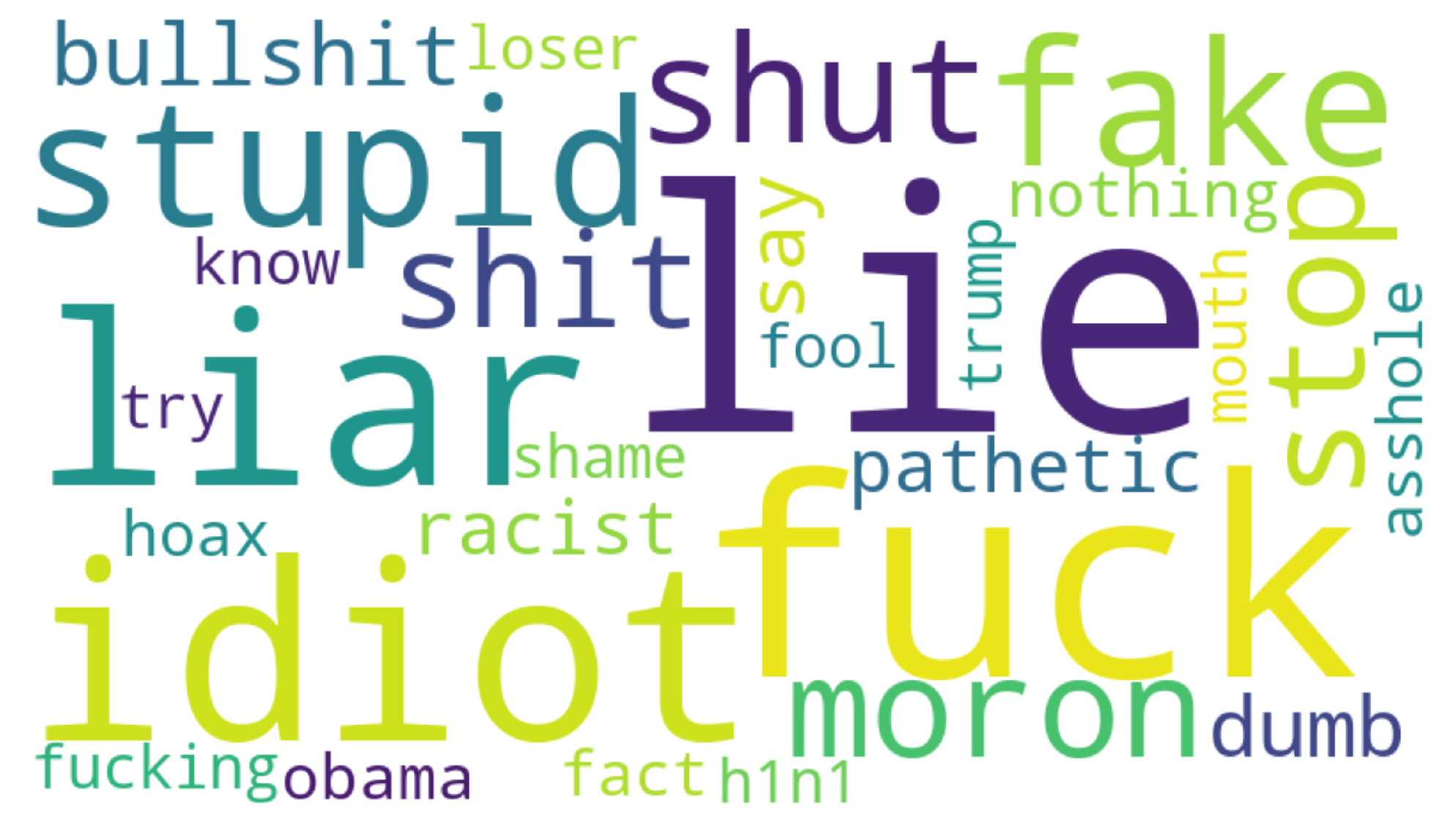}
    \caption{Reply-debunking.}
    \label{fig:1a}
  \end{subfigure}
  \begin{subfigure}[h]{.49\linewidth}
    \centering
    \includegraphics[width=\linewidth]{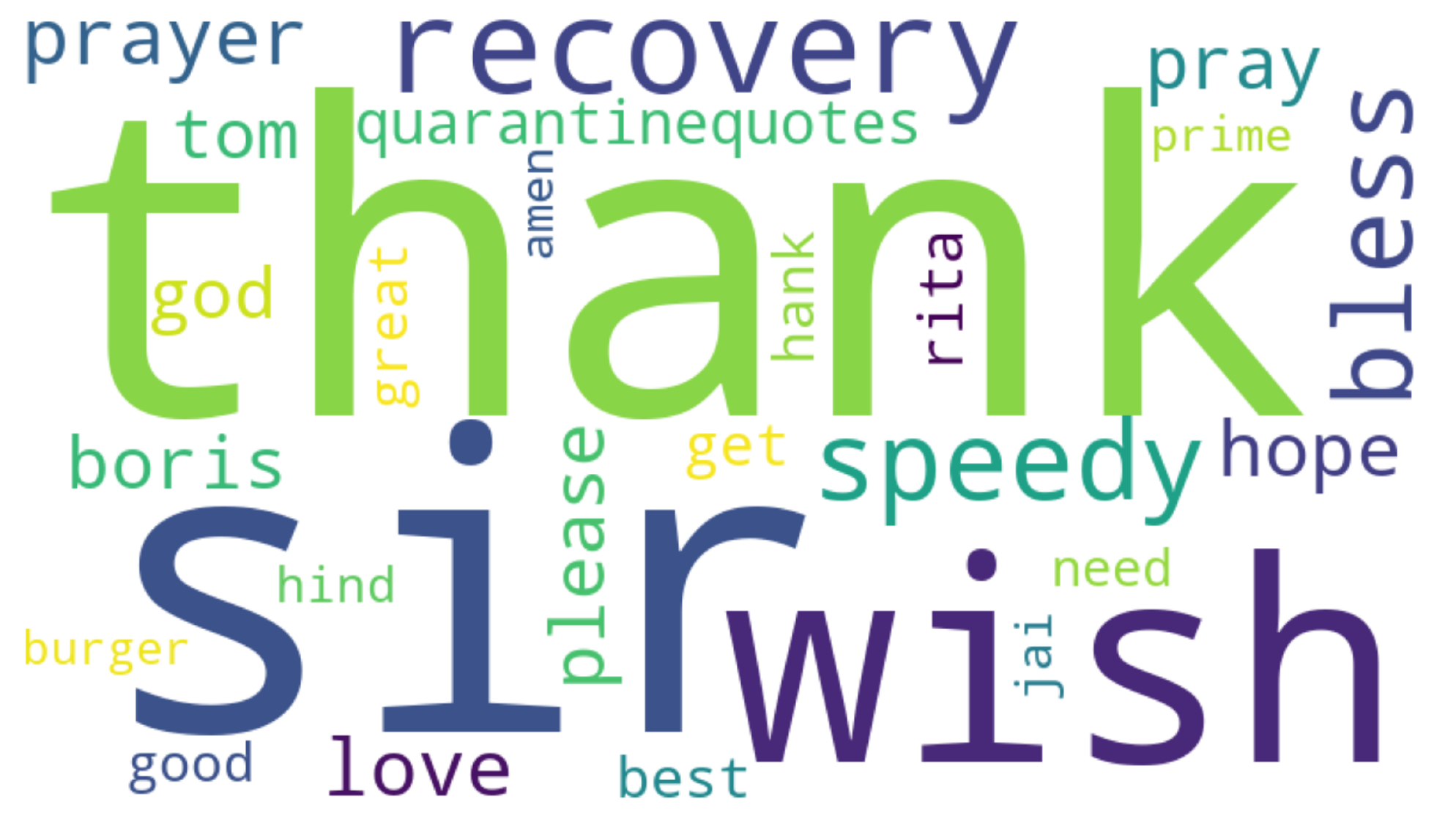}
    \caption{Reply-others.}
    \label{fig:1b}
  \end{subfigure}
  \\
  \begin{subfigure}[h]{.49\linewidth}
    \centering
    \includegraphics[width=\linewidth]{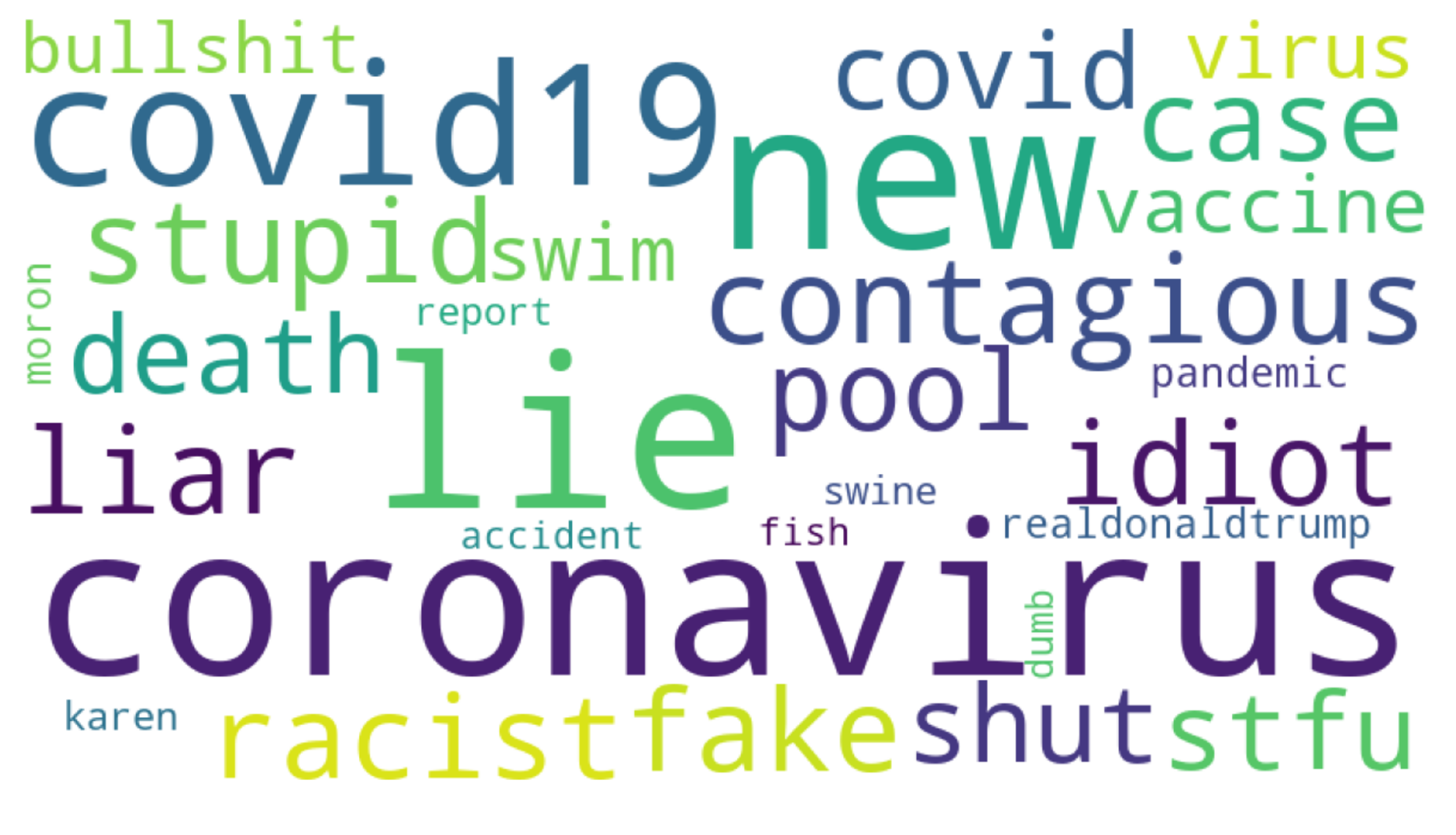}
    \caption{QT-debunking.}
    \label{fig:1c}
  \end{subfigure}
  \begin{subfigure}[h]{.49\linewidth}
    \centering
    \includegraphics[width=\linewidth]{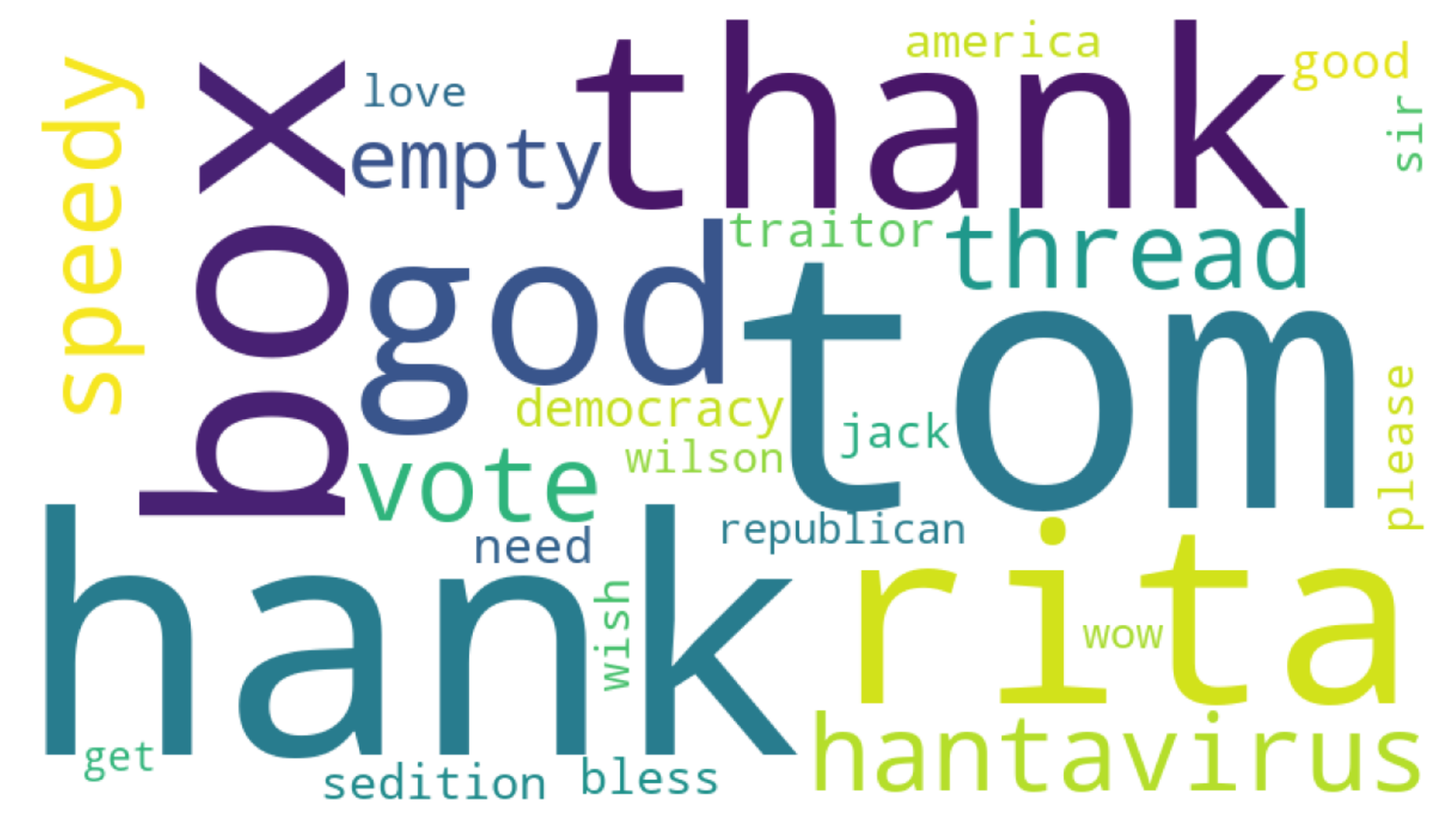}
    \caption{QT-others.}
    \label{fig:1d}
  \end{subfigure}
  \caption{Outstanding words in responses to fake tweets.
  The size of a word in each wordcloud reflects the z-scores by log-odds ratio.
  }
  \label{fig:wordcloud_response} 
\end{figure}

Figure~\ref{fig:zipf} shows its skewed nature; only a small number of fake tweets receive a large number of responses.
In the figure, the relationship between the frequency of responses and debunking a fake tweet gets and its rank shows linearity in log-log space.
Figure~\ref{fig:zipf} also indicates the selection bias of claim-based fake tweets, i.e., manually collected fake tweets.
The response to claim-based fake tweets is skewed slightly away from the linear shape (Figure~\ref{fig:1a},~\ref{fig:1b}), where the frequencies of low-ranked fake tweets are less than the expected values.
This may imply a potential selection bias that more prominent fake tweets are more likely to be collected in previous works.
Alternatively, since the domain-based fake tweets are obtained exhaustively, there is no deviation from the linear shape (Figure~\ref{fig:1c},~\ref{fig:1d}).

\begin{figure}[!htpb]
\centering
\includegraphics[width=\linewidth]{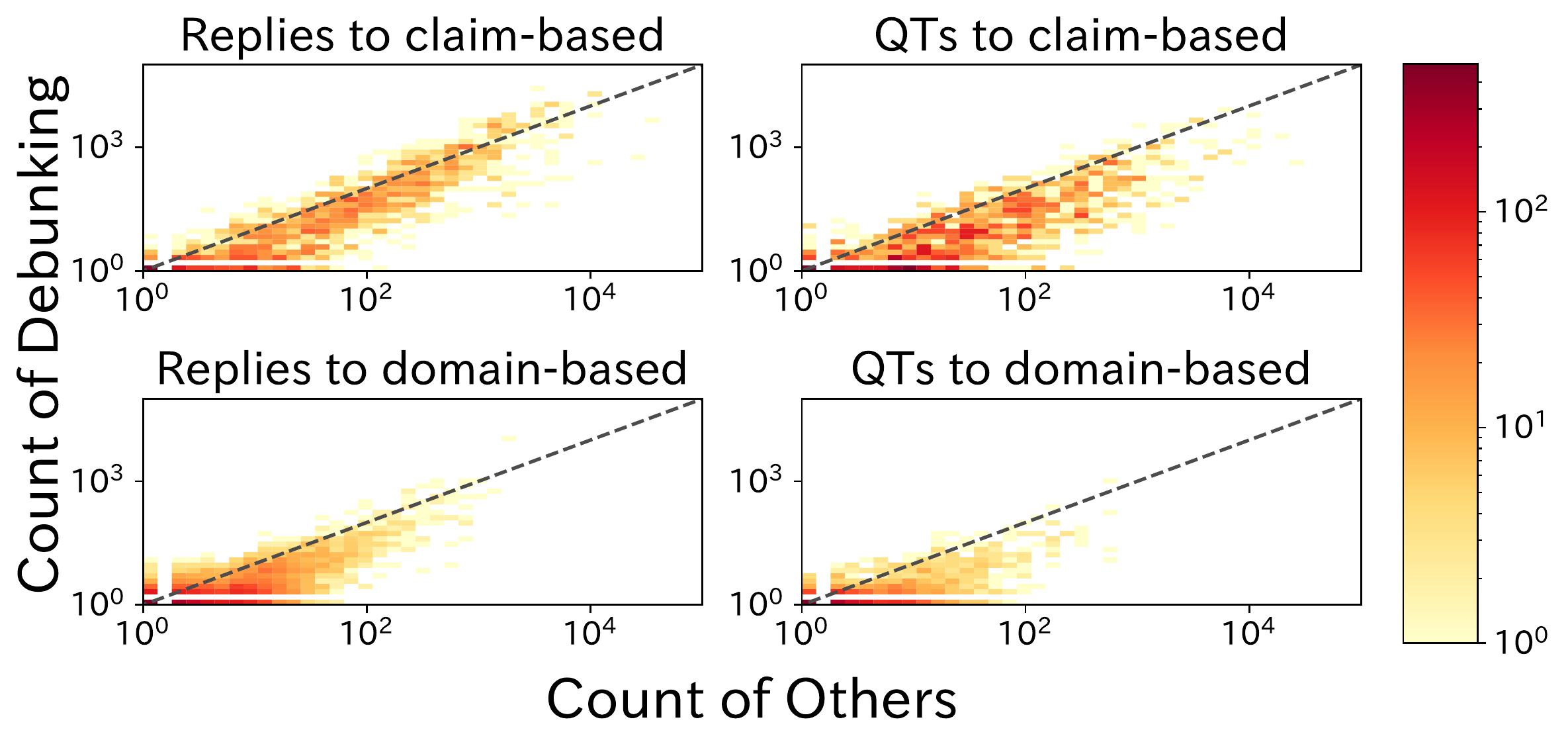}
\caption{Heatmap for fake tweets in terms of the amounts of debunking they get (x-axis) and the amounts of other responses they get.}
\label{fig:debunk_vs_others} 
\end{figure}

\noindent \textbf{Correlation of responses and debunking.}
The more responses a fake tweet receives, the more debunking it receives.
As shown in Figure~\ref{fig:debunk_vs_others}, the number of other responses and debunking received by fake tweets have a linear relationship.
In fact, their correlations are 0.527, 0.586, 0.486, and 0.543 for reply-claim-based, reply-domain-based, QT-claim-based, and QT-domain-based, respectively (p$<$0.001 for all values).

\noindent \textbf{What kind of words in debunking?}
We identify the most representative words of debunking and other responses by using the log-odds ratio~\cite{monroe2008fightin}, which is widely used for comparing multiple corpora~\cite{an2021predicting}.
We aggregate the corpus by aggregating all tweets in a group and comparing them with the tweets of the other group.
We remove terms that appear less than ten times to avoid overemphasis on rare jargon and compute a log-odds ratio of all unigrams.
As the prior, we compute the background word frequency using all tweets in our data collection. 
The unigrams are then ranked by their estimated $z$-scores. 
The result is shown in Figure~\ref{fig:wordcloud_response}.

Debunking contains many words that can be directly used for debunking (i.e., ``lie,'' ``liar,'' ``fake''), regardless of replies and QTs.
Conversely, other responses indicate encouraging words such as ``thank'' and ``love'' in replies and QTs, as well as words related to the conspicuous fake news, such as ``Tom Hanks''~\cite{PolitiFa47:online}, in QTs.

Since we see many offensive words in debunking tweets, we quantify the toxicity of the responses in each group.
We use the Jigsaw's Perspective API\footnote{\label{perspective}\url{https://www.perspectiveapi.com/}} to quantify the toxicity. 
This API measures the toxicity of text on a scale of 0 to 1.
As a result, the median scores are 0.411 for debunking and 0.116 for other responses with $p < 0.0001$ for the Mann-Whitney U test. 

\begin{figure}[!ht]
\centering
  \begin{subfigure}[h]{.49\linewidth}
    \centering
    \includegraphics[width=\linewidth]{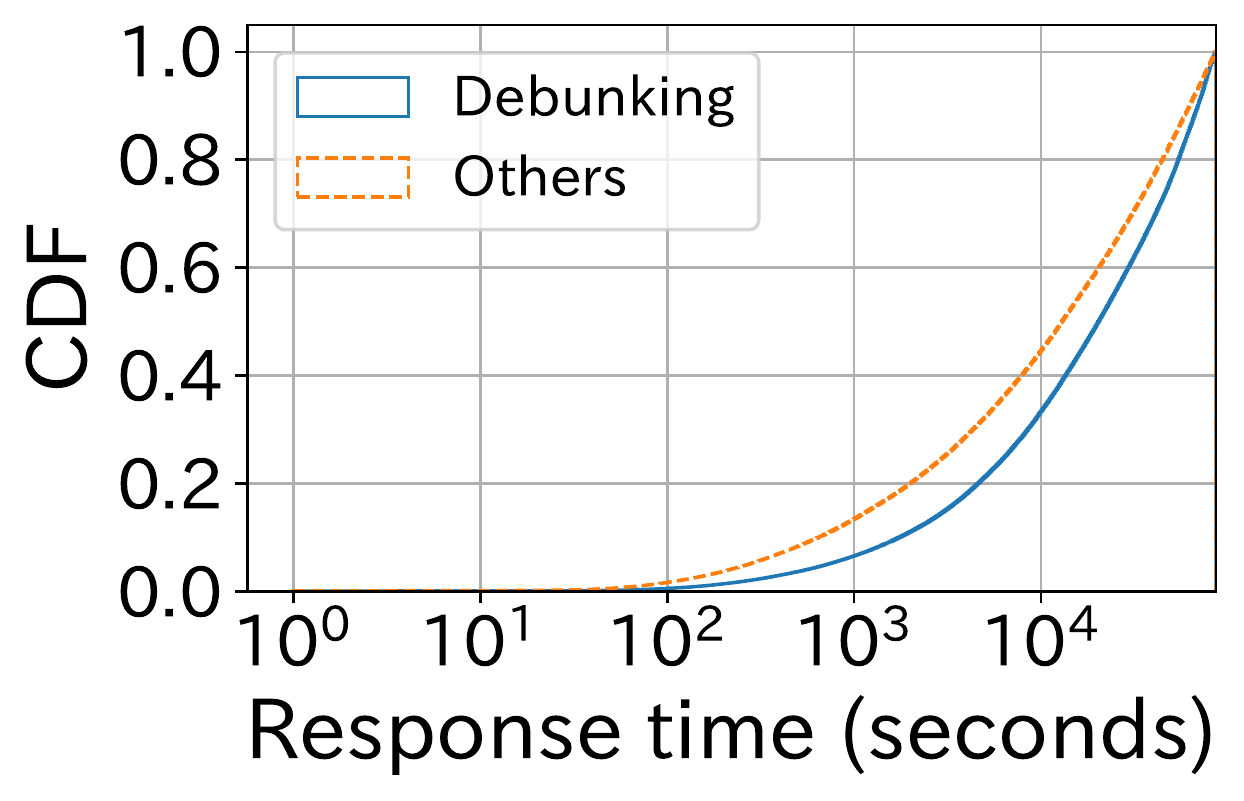}
    \caption{Reply.}
    \label{fig:1a_rt}
  \end{subfigure}
  \begin{subfigure}[h]{.49\linewidth}
    \centering
    \includegraphics[width=\linewidth]{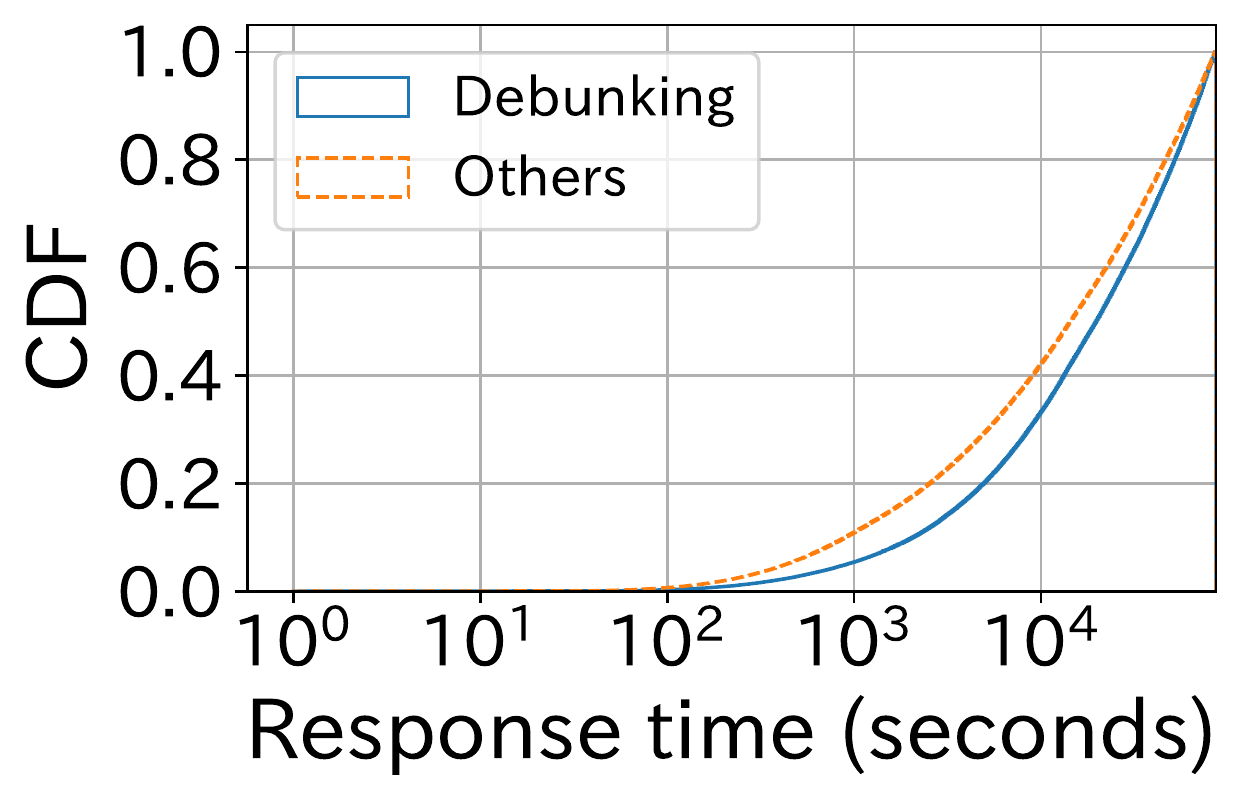}
    \caption{QT.}
    \label{fig:1b_rt}
  \end{subfigure}
  \caption{CDF of interval time between the fake tweets and their responses.}
  \label{fig:response_time} 
\end{figure}

\noindent \textbf{How quickly  are fake tweets debunked?}
We also examine the timing of debunking.
Figure~\ref{fig:response_time} shows the CDF of the interval time between the responses and the targeted fake tweets.
We can see that debunking occurs later than other responses in Reply and QT.
For example, the median intervals in other responses were 13,024 seconds in reply and 14,429 seconds in QT, while for debunking, they were 20,438 seconds in reply and 19,820 seconds in QT.
Other responses are faster than debunking at $p < 0.0001$ with the Mann-Whitney U test. 
Debunking is slower than other responses probably because debunking is an action that generally requires a lot of thought and research, and therefore takes more time on average than other responses.

\section{RQ1: What are the characteristics of fake tweets that tend to be debunked?}

We investigate the potential reasons and motivations for users to debunk fake tweets. 
In particular, we focus on three different perspectives of fake tweets: who is sharing the tweet (\textit{User features}), what the tweet is about (\textit{Content features}), and how people react to it (\textit{Engagement features}).
We then conduct regression analysis to examine which features have the strongest predictive power to detect which fake tweets are likely to be debunked.
\bm{Specifically, we predict whether or not fake tweets get at least one debunking response by using the logistic regression model, and analyze the regression coefficiency of the features.
We conduct the prediction and analysis for reply and QT separately.}

\subsection{Regression Features}
We select the following features that are expected to be relevant to the conditions of occurrence of spontaneous debunking based on related works.

\noindent \textbf{User features.}
We use (1) followers count, i.e., how conspicuous the users are, (2) length of bio, i.e., user's willingness to appeal, and (3) verification by Twitter, i.e., user's publicity.
Verification by Twitter is a binary measure of whether or not Twitter has authorized an account\footnote{\url{https://help.twitter.com/en/managing-your-account/about-twitter-verified-accounts}}.

\noindent \textbf{Content features.} 
We consider topics and styles of fake tweets as content features. 

\textit{Topic features.} 
For topic extraction, we use a biterm topic model~\cite{yan2013biterm}, which is known to work better for short sentences than other widely-used topic models such as LDA.
We choose seven as the number of topics by comparing the perplexity scores~\cite{zhao2015heuristic}.
We assign each tweet to one of the seven topics as dummy variables once we build the topic model.
We determined that assigning one topic with the highest probability of belonging to one tweet is sufficient.
The seven topics, their representative words, and the proportions of tweets for each topic are summarized in Table~\ref{table:topics}.
Early emergence and spread of COVID-19 in other countries (e.g., China, Russia, and Iran~\cite{Russiasc66:online}) and politics are the two most prevalent topics among fake tweets during COVID-19. 
We note that we use ``Cases/Deaths'' as a reference when creating dummy variables of topics for building the regression model to avoid the ``dummy variable trap''~\cite{gujarati1970use}. 

\bm{In addition, to account for potential differences between the two types of datasets, we also add a binary feature, whether an original fake tweet is claim-based or domain-based (Type of fake tweet: 1 for claim-based; 0 for domain-based), to the model.}

\begin{table}[!ht]
\centering
\small
\begin{tabular}{p{2.3cm}p{0.7cm}p{4.0cm}}
\hline
Topic labels             & Ratio & Representative words                                                                \\ \hline
Control measures         & 2.6\%       & ban, governor, lockdown, order, michigan, whitmer, house, que, force, task          \\ 
Political figures        & 21.6\%      & biden, watch, video, mask, cuomo, joe, die, american,   lockdown, donald            \\
Status around the world  & 25.5\%      & lockdown, china, outbreak, due, test, case, crisis, russia,   spread, iran          \\
Cases/Deaths         & 14.3\%      & death, case, number, cdc, rate, toll, day, patient, state,   study                  \\
Vaccine                  & 12.4\%      & test, vaccine, positive, gate, bill, fight, covid, corona,   treatment, study       \\
Wuhan lab                & 12.9\%      & china, wuhan, lab, chinese, claim, time, hoax, scientist,   doctor, video           \\
Players against COVID-19 & 10.7\%      & bill, pelosi, democrat, china, relief, house, illegal,   american, stimulus, crisis \\ \hline
\end{tabular}
  \caption{Topics of fake tweets by biterm topic model. 
  The ratio is the percentage of tweet count assigned in each topic per total.
  Representative words are the top 10 words with the highest probability of occurring in each topic.
  }
  \label{table:topics} 
\end{table}

\textit{Linguistic features.} 
We use features that capture the linguistic style of fake tweets, such as (1) length of the text, i.e., the volume of information of tweets; (2) the existence of a URL, i.e., the existence of evidence; and (3) sentiment, i.e., impressions to readers.
For the sentiment, we use a pre-trained model of~\citet{barbieri2020tweeteval} and use the positive and negative values.

\textbf{Engagement features.}
We use the retweet count as the feature of attention to fake tweets.
The count of favorites is alternative, but they are highly correlated and produce multicollinearity; thus, we only choose retweet count.
\bm{In addition, we consider the numbers of replies and QTs as features to account for the general likeliness of receiving replies and QTs.}

\subsection{Regression Results}
Table~\ref{table:regression} shows the results of the logistic regression. 
The p-values are computed using two-tailed z-tests. 
All variance inflation factors (VIFs)~\cite{o2007caution} are less than three, indicating that multicollinearity is negligible. 
We conduct log transform to some features when needed, indicated as (log) in Table~\ref{table:regression}.

Among the user features, followers count has positive associations in replies and QTs, which means tweets from conspicuous accounts are more likely to get debunks.
Among the topic features, \bm{the topic of Wuhan lab has positive and significant coefficients both in reply and QT.
The topic of political figures and players against COVID-19 has positive and significant coefficients in QT.
In contrast, fake tweets about status around the world and control measures against COVID-19 are less likely to get debunked both in reply and QT. 
For the other topics, the coefficients tend to be greater than 0, although some of them are not significant.}
\bm{Also, claim-based fake tweets are less likely to get debunked compared to domain-based fake tweets (the type of fake tweet has significant negative coefficients).}
Among the linguistic features, fake tweets with URLs and negative fake tweets are more likely to get debunked in both reply and QT. 
\bm{Lastly, reply count shows a positive association for both reply and QT, and QT count has a negative value in reply and positive value in QT.}

\begin{table}[!ht]
\centering
\small
\begin{tabular}{p{2.4cm}lccc}
\hline
                         & Reply & QT     & Mean & Std  \\ \hline
(Intercept)              & -8.51${}^{***}$ & -10.84${}^{***}$ & -    & -    \\ \hline
Followers   count (log)  & 0.08${}^{***}$  & 0.23${}^{***}$   & 7.75 & 2.68 \\
Length of bio (log)      & 0.03${}^{*}$   & 0.13${}^{***}$   & 4.03 & 1.58 \\
Verification             & 0.03${}^{}$   & -0.28${}^{***}$  & 0.11 & 0.32 \\ \hline
Control measures         & -2.02${}^{***}$ & -3.57${}^{***}$  & 0.03 & 0.16 \\
Political figures        & 0.02${}^{}$   & 0.44${}^{***}$   & 0.22 & 0.41 \\
Status around the world  & -0.60${}^{***}$ & -0.39${}^{***}$  & 0.26 & 0.44 \\
Vaccine                  & -0.07${}^{}$  & 0.18${}^{*}$   & 0.12 & 0.33 \\
Wuhan lab                & 0.26${}^{***}$  & 0.36${}^{***}$   & 0.13 & 0.34 \\
Players against COVID-19 & 0.00${}^{}$   & 0.56${}^{***}$   & 0.11 & 0.31 \\ \hline
\bm{Type of fake tweet}             & -1.67${}^{***}$ & -1.88${}^{***}$  & 0.03 & 0.17 \\ \hline
Length of text (log)     & 0.04${}^{}$   & 0.20${}^{**}$   & 5.20 & 0.35 \\
URL                      & 2.43${}^{***}$  & 1.09${}^{***}$   & 1.00 & 0.07 \\
Positive                 & 0.12${}^{}$   & -0.39${}^{**}$  & 0.09 & 0.17 \\
Negative                 & 0.59${}^{***}$  & 0.68${}^{***}$   & 0.43 & 0.29 \\ \hline
Retweet count (log)      & -0.15${}^{***}$ & 0.03${}^{}$   & 0.44 & 0.93 \\
\bm{Reply count (log)}        & 3.95${}^{***}$  & 0.10${}^{**}$   & 0.13 & 0.54 \\
\bm{QT count (log)}           & -1.21${}^{***}$ & 2.53${}^{***}$   & 0.07 & 0.42 \\ \hline
Pseudo R-squ.            & 0.57  & 0.55   & -    & -    \\ \hline

\multicolumn{5}{l}{
  Significance codes:
  ***$p<0.001$, 
  **$p<0.01$, 
  *$p<0.05$
  }
\end{tabular}
  \caption{Results of the regression analysis predicting whether or not a fake tweet gets debunked (number of
    fake tweets is N = 343,549). Each number indicates the regression coefficients.
  }
  \label{table:regression} 
\end{table}

In summary, as expected, conspicuous accounts are more likely to receive debunking (follower count), \bm{and tweets that get more replies tend to get debunked in reply, as does QT.} 
Also, tweets expressing negative emotions and containing URLs tend to get debunked.
After controlling for these factors and focusing on topics, we find that \bm{debunking is more common for Wuhan lab in both reply and QT. 
Politics and players against COVID-19 get more debunking in QT.} 
In contrast, there is less debunking to the topics such as the status of infection, measures around the world, and the countermeasures against COVID-19 itself.

\section{RQ2: What are the characteristics of spontaneous debunkers?}

In this section, we characterize debunkers by profile descriptions and networks. 

We first examine how active users are in debunking fake tweets. 
Among all respondents, almost half of the users have debunked at least once (Debunking $\geq$ 1) by replies (340k (48.5\%)), and a quarter of the users have debunked by QTs (97k (25.9\%)). 
Those who have debunked three or more times (Debunking $\geq$ 3) are 46k (6.6\%) with replies and 5.8k (1.6\%) with QTs,
indicating that it is not so common to debunk multiple times, but there is a considerable number of users who frequently debunk fake tweets. 

\subsection{Who are debunkers?} 

\subsubsection{Difference between debunkers and non-debunkers.}
To better understand debunkers, we conduct a comparative analysis between debunkers and non-debunkers.
For comparison, we define debunkers as those who have debunked more than 3 times and non-debunkers as those who have responded to fake tweets more than 3 times but have never debunked them.
This criterion is to reduce the impact of the error of our debunking detection model on the analysis.

\begin{figure}[!ht]
\centering
  \begin{subfigure}[h]{.49\linewidth}
    \centering
    \includegraphics[width=\linewidth]{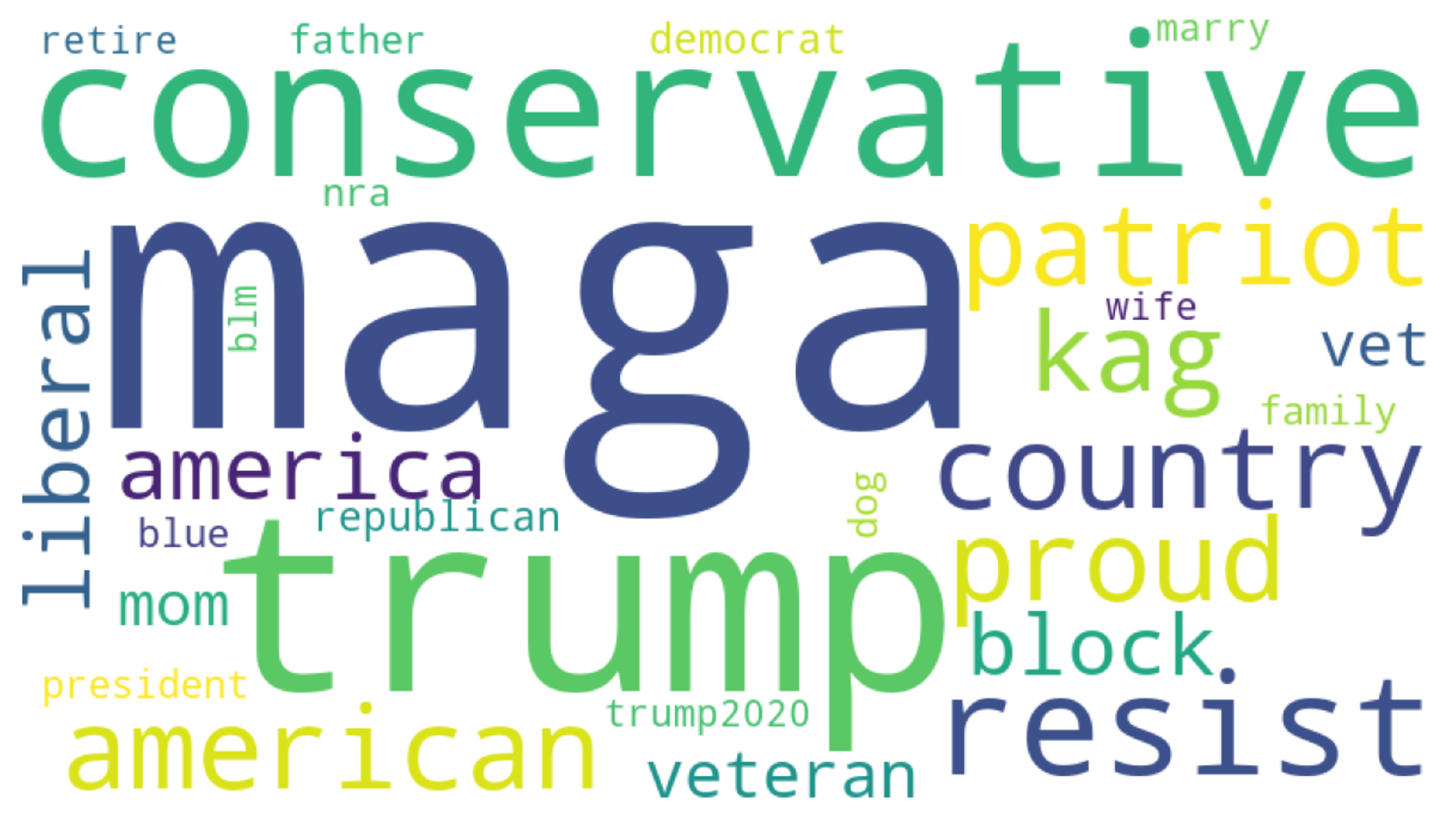}
    \caption{Debunkers.}
    \label{fig:1a_wb}
  \end{subfigure}
  \begin{subfigure}[h]{.49\linewidth}
    \centering
    \includegraphics[width=\linewidth]{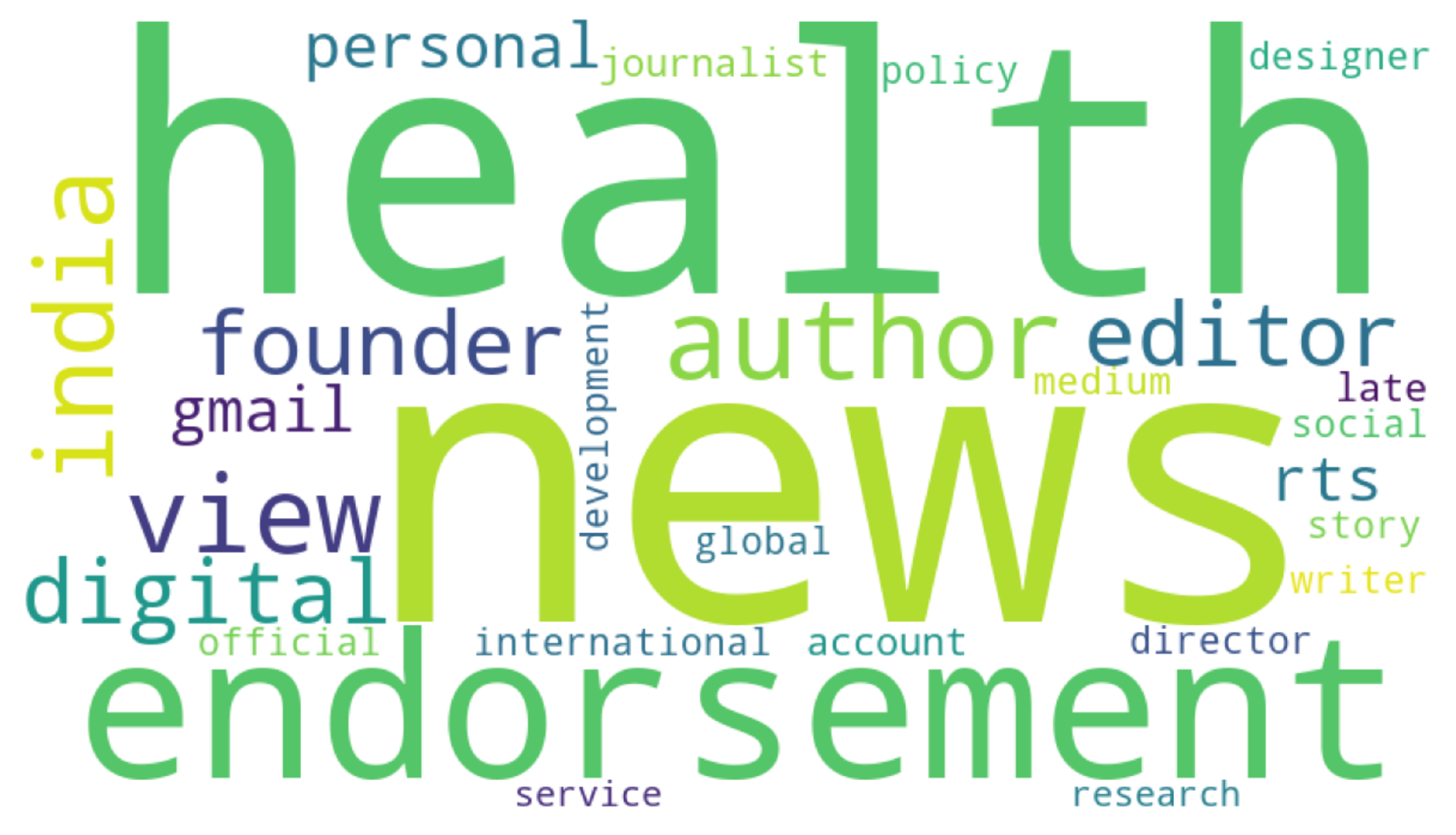}
    \caption{Non debunkers}
    \label{fig:1b_wb}
  \end{subfigure}
  \caption{Outstanding words in bios in debunkers and non-debunkers.
  The size of words in each wordcloud is along with the z-scores by log-odds ratio.
  }
  \label{fig:wordcloud_debunkers} 
\end{figure}

We examine the difference in bio descriptions between debunkers and non-debunkers.
We calculate the log-odds ratio of words in bios of debunkers and non-debunkers~\cite{monroe2008fightin}. We use our COVID-19-related tweets as a background corpus. Figure~\ref{fig:wordcloud_debunkers} shows the representative words of debunkers' and non-debunkers' bios.
Interestingly, we see many political words in debunkers' bios. For example, words relating to conservatives, such as ``maga,'' ``conservative,'' and ``trump,'' and those relating to liberals, such as ``blue,'' ``liberal,'' and ``democrat'' are observed.  
In other words, debunkers tend to be highly partisan accounts.
Conversely, non-debunkers' representative words are far less political and include words like includes ``health,'' ``news,'' and ``endorsement'' (e.g., ``RTs are not endorsements'').

\subsubsection{Types of debunkers.}
Next, we identify types of debunkers by applying the biterm topic model to their bios.
We then examine the average debunking ratio for each group obtained by the topic model.
Table~\ref{table:user_topic} shows that the highest debunking ratio is found in the conservative group, followed by the liberal group, and then those without bios.
In contrast, the users who have information about Business/Politics/Health in their bio have the lowest debunking ratio.
The difference between these groups is significant at p $<$ 0.001 for all pairs of Mann-Whitney U tests with Bonferroni correction.
Business/Politics/Health is the only group that has the word ``science'' in the top 20, although it is not listed in the table, so it is likely that these users are relatively interested in science, but their debunking ratio is low.

\begin{table}[!ht]
\centering
\scalebox{0.92}{
\begin{tabular}{p{1.6cm}p{0.7cm}p{3.8cm}p{0.8cm}}
\hline
Topic labels             & User ratio & Representative words                                                                & Debunking ratio \\ \hline
Conservative             & 16.3\%     & trump, proud, god, conservative, maga, husband, father, american, family,   country & 0.267                   \\
Business /Politics /Health & 22.4\%     & view, writer, former, business, politics, news, health, opinion,   director, social & 0.166                   \\
Liberal                  & 24.0\%     & mom, resist, blm, wife, dog, animal, life, proud, liberal, mother                   & 0.230                   \\
Spiritual                & 7.7\%      & life, people, good, world, live, truth, work, try, god, take                        & 0.205                   \\
Hobby                    & 15.8\%     & fan, sport, music, game, husband, football, dad, father, movie,   enthusiast        & 0.192                   \\
No bio                   & 13.8\%     & -                                                                                   & 0.223                   \\ \hline
\end{tabular}
}
\caption{Topics in the bios of respondents. The debunking ratio is the average ratio of debunking tweets among their responses to fake tweets.}
\label{table:user_topic} 
\end{table}

\subsection{How are debunkers connected?}
Finally, we examine social connections between debunkers to understand how far or close they locate on social media.
We take the top 1,000 users with the most debunkings in replies and QTs each (1,644 unique users), and get all of their followers. We use Louvain clustering~\cite{blondel2008fast} to identify clusters.

\begin{figure}[!ht]
\centering
\includegraphics[width=\linewidth]{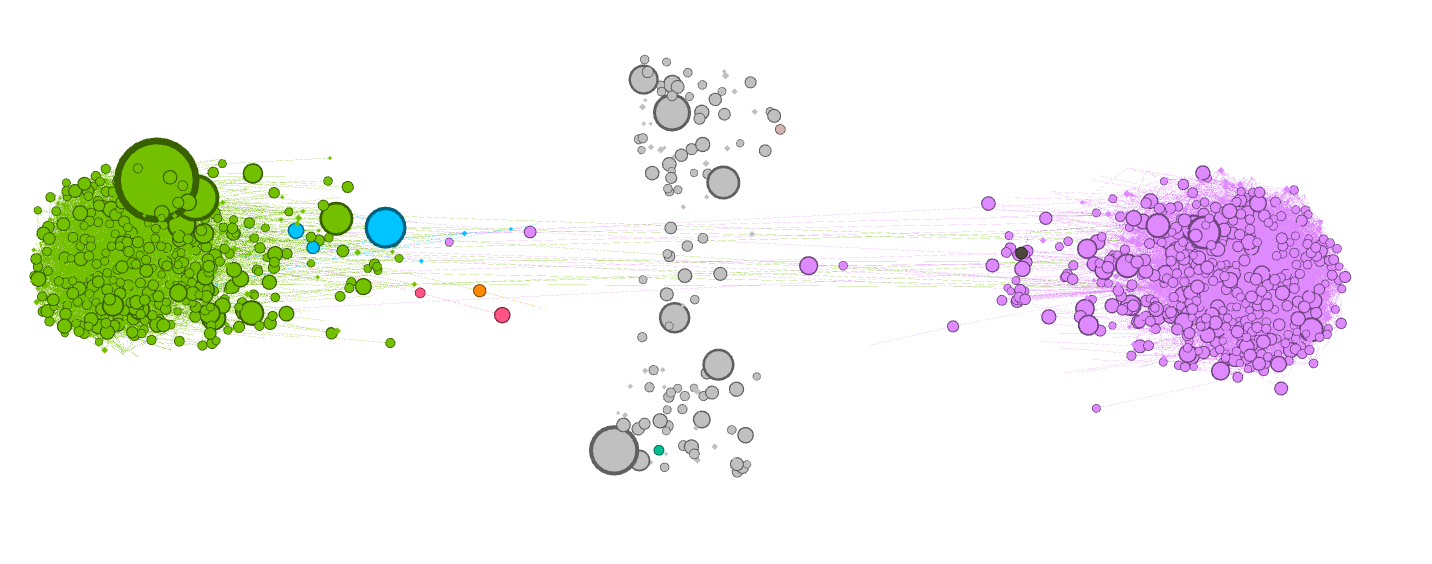}
\caption{Follower network of top debunkers. Nodes indicate the users who conduct the debunking the most frequently (top 1,000 in reply and QT, 1,664 in total without duplicates). The size of nodes corresponds to the frequency of debunking. Edges indicate the follower relationship.}
\label{fig:network} 
\end{figure}

Figure~\ref{fig:network} shows the follower network with identified clusters.
We observe two large, separated clusters, which are slightly connected with each other. This indicates a clear echo-chamber/polarization phenomena~\cite{barbera2015tweeting}.
By looking into the bios of each cluster, we found that purple is conservative (50.06\%), and green is liberal (41.24\%).
\bm{Other than these two large clusters, the rest of the users in the center of Figure~\ref{fig:network} are all isolated, not even connected with the two large clusters.}
Overall, the partisan debunkers are highly connected, and they can see the debunking behavior of each other, which may lower the hurdle of debunking behavior.

\section{Discussion and Conclusion}
\subsection{Main findings}
We have shown several remarkable findings in spontaneous debunking on Twitter.
First, we confirm that many fake tweets are left undebunked
and debunking behavior is generally slower than other responses.
If we expect spontaneous debunkers to act as social sensors~\cite{liu2015social}, we need to design an environment that incentivizes them to debunk more and faster.
Second, the topics of debunked fake tweets are unevenly distributed. 
\bm{Debunkings tend to be directed at fake tweets about Wuhan lab, seeming to be the most apparent conspiracy theory.
Also, fake tweets about political figures and players against COVID-19 tend to get debunked partially.} 
Instead, there was relatively little debunking to fake tweets about the global infection/control situation and infection control measurement itself (e.g., lockdown, masks).
These may be topics that fact-checkers should focus on, as understanding the status of infections and countermeasures is vital information to encourage people to act rationally.
Third, we found that the most frequent debunkers are partisan and connected well in each group.
It seems political motivation drives debunking behaviors. 
In other words, we may consider enlisting the help of their debunking behavior for politically relevant topics.
\bm{However, according to~\citet{allen2020scaling},  politically balanced people are more accurate in debunking; thus, the debunking of partisan users has to be used with caution.}
Then again, it may be necessary to encourage users with different interests to do the debunking for other topics.
For example, researchers were requested to cooperate in countering fake news even before COVID-19~\cite{lazer2018science}.

\subsection{Limitations and future works}
\noindent \textbf{Credibility of social correction.}
\bm{Even though a ``wisdom of the crowds'' is highly correlated with professional fact-checking, as~\citeauthor{yasseri2021can} pointed out, there is a risk in relying entirely on the social correction. 
In particular,~\cite{allen2020scaling} reported that when the partisanship of debunkers becomes high, the correctness of debunking lessens.
It is challenging to identify  meaningful information from social correction for fact-checking.}

Alternatively, it may be possible to examine the differences in debunking among different types of debunkers.
For example, this study showed that many of the top debunkers are partisan, and it could be possible that what is true for one side may be seen as a fake by the other side.
In such a situation, it would be interesting to find out which posts were judged as false by both parties.

\noindent \textbf{Definition of debunking.}
\bm{Our working definition of debunking is an action to inform authors of posts as inaccurate.
We use this broad definition in order to capture the overall tendency of possible debunking behaviors after separating all responses into debunking and other responses.
Our analysis then found multiple styles of debunking, including ``pointing out'' and ``insulting.''
We have to note that insulting is not necessarily considered as debunking in some cases, as such behavior is often driven by partisan motivation in political information. 
A detailed categorization of debunking and characterization of different categories can be considered in a future study.}
Also, while this study focuses on debunking, focusing on supporting responses may help to find fake tweets that people tend to accept.
In this case, we would find blind spots to improve digital media literacy~\cite{THELANCET20221}.

\noindent \bm{\textbf{Classification accuracy}
This study is highly dependent on the quality of the classifier. 
Therefore, we built a model with high accuracy to conduct the subsequent analyses and the performance of the model is comparable with an existing study~\cite{he2021racism}.}
\bm{To demonstrate the generality of our fine-tuned model, we also conducted a manual evaluation of the prediction results (in $\S$4.2).}

\bm{Moreover, we further conducted an error analysis to understand how to improve our model.
We hypothesized that it would also be difficult for the annotators to make an accurate decision for the tweets the model failed to predict.
We analyzed the relationship between the disagreement of the annotators' judgments and the incorrectness of the model prediction.
Specifically, we calculated the F1 score with respect to a binary variable of whether the three annotators agreed or disagreed and a binary variable of whether the predicted result was correct or not, and found a certain positive correlation: 0.63 for replies and 0.62 for QTs.
The contents that are difficult for people to judge are also difficult for the classifier.
However, there are various patterns in missed tweets, so further research may be needed.}

\bm{Lastly, it would be better to achieve even greater accuracy in the classification of debunking behavior. Therefore, future research includes further improvement of accuracy.
When it comes to the balance of recall and accuracy, both the certainty of the predicted result and the target coverage are important, and it is not easy to prioritize one over the other. However, we note that precision would be more important considering the confidence in the subsequent analyses.}

\noindent \textbf{Bias in datasets}
The domain-based approach uses keyword-based tweet collection to gather COVID-19-related tweets, but not all COVID19-related tweets may have been collected, such as the name of emerging variants. However, since this is a trend analysis using large-scale data, some lack of keywords can be considered to have a minor impact on the analysis.

\bm{In addition, in this study, we assumed all information from suspicious domains as false information. While this is a common assumption used to study fake-news detection and spread~\cite{baly-etal-2020-written}, not all information is completely false, even if the domain is suspicious. In particular, factual information, such as infection status, is less likely to be false, which may have affected the regression analysis results. A deeper investigation may be required in the future.}

\bm{In terms of the result of our  regression analysis, we found that people tend to debunk domain-based fake news more than claim-based fake news, which is consistent between RT and QT. 
This indicates that the debunking of claim-based fake tweets is less likely to occur than domain-based fake tweets. 
This may imply that ``suspicious domains'' can be an easy indicator of fake information, and thus,  domain-based fake tweets are more easily identified and debunked. This result highlights that online space needs better support for recognizing claim-based fake tweets.}

\noindent  \bm{\textbf{Application to other topics}
This work introduces a framework to analyze debunking behaviors. Since the analysis procedure of this study is not topic-dependent (i.e., we can collect and analyze fake tweets and responses on an arbitrary topic), it can also be applied to other topics.
In particular, since some debunking behavior is not dependent on the content of the fake tweet (e.g., debunking such as "This is fake" can exist in any topic), it is possible that our  debunking classifier can be applied to other fake news in different domains. 
However, some debunking behaviors are indeed content-dependent, and we cannot make accurate speculation at this time. Transfer learning of debunking classification sounds like a very interesting topic and we will leave it as a future research topic.}

\subsection{Ethical Considerations}    
We pay the utmost attention to the privacy of individuals in this study. 
We did not include personal names or account names in our analysis. Moreover, in the example figures, we blur user identity-related features (name, photo, and user id) to maintain anonymity.
Lastly, for sharing our tweet data, we will publish only a list of tweet IDs, without any text or information, according to Twitter's guidelines.

\section*{Acknowledgement}
This work was supported by JST, CREST Grant Number JPMJCR20D3, Japan.

\bibliography{main}

\end{document}